# An efficient scheme to optimize the superconducting levitation via genetic algorithm

Chang-Qing Ye[1,2,3], Guang-Tong Ma[1,2,a)], Xing-Tian Li[1,2,4], Han Zhang[1,2,4], Peng-Bo Zhou[1,2], Chen Yang[1,2,4], and Jia-Su Wang[2]

[1]*State Key Laboratory of Traction Power, Southwest Jiaotong University, Chengdu, Sichuan 610031, China*
[2]*Applied Superconductivity Laboratory, Southwest Jiaotong University, Chengdu, Sichuan 610031, China*
[3]*School of Mechanical Engineering, Southwest Jiaotong University, Chengdu, Sichuan 610031, China*
[4]*School of Electrical Engineering, Southwest Jiaotong University, Chengdu, Sichuan 610031, China*

## ABSTRACT

The superconducting levitation consisting of high-temperature superconductors (HTSs) and permanent magnet guideway (PMG) is deemed promising technique for the advancement of the maglev transit. To improve the cost-efficiency and thus reduce the investment of this superconducting levitation transit, the optimization of the PMG is the most critical issue of practical interest since it serves as the continuous rail to generate the magnetic field by the rare-earth magnets. By the use of a generalized vector potential within the quasistatic approximation as the state variable to mathematically describe the HTS as well as the surrounding medium, an efficient scheme for optimizing the superconducting levitation has been developed with the genetic algorithm as a strategy to perform the global search of the PMG. This scheme directly describes the HTS element without simplification of its *intractable* nonlinearity of constitutive law, which renders this study stand out from the existing efforts. The testing of the proposed scheme on a typical optimization of the superconducting levitation has proven its robustness and efficiency, i.e., the time cost is merely 3.6 hours with 3000 individuals evaluated on a moderate desktop. Taking a HTS over the Halbach-derived PMG as a practice, a set of case studies were carried out to understand how the working condition, geometrical and material characteristics of the HTS affect its maximum levitation force achievable at different constraints of the cross-section of the PMG. The findings attained by the case studies, being *inaccessible* from the experiments, are aimed to provide useful implications for the optimization of a superconducting levitation system for the transit and analogous purposes.

[a)]Author to whom correspondence should be addressed. Electronic mail: gtma@home.swjtu.edu.cn.

## I. INTRODUCTION

Physical levitation that uses the magnetic force instead of the contact force to balance the gravity has been explored for century in a wide range of industrial applications–particularly in the field of rail transit,[1-5] whose operating speed could be considerably upgraded without the mechanical contact between the vehicle and rail. One of the promising techniques for realizing such kind of levitation is to immerse the high-temperature superconductors (HTSs) in a nonuniform magnetic field generated by a specially designed permanent magnet guideway (PMG).[4,5] The superconducting elements in this situation work in an inductive mode and their electromagnetic interaction with the external field could provide an inherently stable levitation.[2] The research and development of this superconducting levitation for the rail transit purpose has been continually advanced for over a decade[6] and currently, demonstration of such technique towards a full-scale level is being carried out in different groups.[5,7]

To improve the cost-efficiency and thus reduce the investment of the named superconducting levitation transit, the optimization of the PMG is thought to be most critical from a practical point of view since it serves as the continuous rail to generate the magnetic field by the rare-earth magnets.[4] The existing efforts made to this important issue by means of either simulation[8-17] or measurement[18-21] have clarified that, the configurations of the PMG derived from the Halbach array[22] are superior in virtue of their outstanding function to concentrate the magnetic field in the desired region where the superconducting elements are levitated. The geometrical effect of the Halbach-derived PMGs on the performance of the superconducting levitation has been widely studied, among which the dimensional ratio of the magnet elements in the PMG to achieve an optimized levitation system has been suggested.[8,9,11,15,17,18]

These abovementioned achievements, to some extent, lead to a cost-efficient design of the superconducting levitation. However, to make a global search and thus study this important issue in a higher degree, an intelligent and efficient scheme of optimization in which the geometrical parameters of the PMG could be randomly generated should be developed. Since its importance, recently published papers have been devoted to this aspect. The initial work in global optimization coupled with the genetic algorithm was done by Motta *et al*., who suppose that the superconducting elements are perfect diamagnet with null permeability of the frozen-image model.[23] This scheme uses the analogous methodology to adapt the ANSYS software and the simulated results of levitation force are said to be overestimated. More recently, Hekmati applied simulated annealing method to optimize a simplified levitation system, which models the superconducting elements by an analytical model with Bean's model of the critical-state and are far from the practical requirement.[24]

Described in this work is an efficient and robust scheme for intelligently optimizing the superconducting levitation system. It uses a self-developed 2-D nonlinear model to simulate the magnetic forces of HTS over a PMG[25,26] and the genetic algorithm as a strategy to perform the global search. The finite-element technique is employed to discretize the spatial domain and the resultant nonlinear-large-sparse matrix equation is treated by the Jacobian-free Newton-Krylov approach,[27] which is shown to be fast and robust in our practice.[25] With this scheme we obtain the quantitative results that help delineate the relationships between the levitation performance and the geometrical and material characteristics of the HTS and the PMG.

Calling upon the quasistatic approximation of a generalized vector potential, in Sec. II we define the theoretical model of the superconducting levitation and set out the general framework which lays the basis of the numerical simulation. The basic knowledge of the genetic algorithm in this consideration as well as how to couple it with the numerical method in our scheme is briefly introduced in Sec. III. Making recourse to these, in Sec. IV we calculate a set of representative examples and analyze the computational performance to reveal the excellent efficiency of the proposed scheme in optimizing the superconducting levitation. Using the developed optimization scheme, we performed case studies in Sec. V to investigate the geometrical and material effect on the optimum outcome of levitation force in the purpose of suggesting implications for designing the high cost-efficient PMG. The conclusion is presented in Sec. VI to summarize and highlight this work.

## II. NUMERICAL MODEL

### A. Theoretical foundations

This section describes the mathematical formulations we use to govern the electromagnetic behavior of HTS in the presence of a nonuniform magnetic field. We simplified the actual 3-D problem to be a 2-D one in view of the practical case that the PMG usually extends infinitely. This simplified 2-D model, with less complexity compared to the 3-D model,[28,29] has an economic applicability in simulating and solving the HTS levitation problem with translational symmetry, especially for the purpose of optimization where thousands of commands of the electromagnetic module is generally required. Fig. 1 presents the geometrical configuration of the studied superconducting levitation system with a HTS levitated over a Halbach-derived PMG.

According to the geometrical coordinate shown in Fig. 1, we will briefly introduce the 2-D theoretical model of HTS subject to magnetic stimulation. This model was established by defining a generalized magnetic vector potential as follows,

$$A'_{sc,x} = A_{sc,x} + \int_0^{t_n} C(t)dt. \tag{1}$$

where $A_{sc,x}$ represents the vector potential induced by the supercurrent in the HTS and $C(t)$ is a time-dependent variable to describe the gradient of electric scalar potential at an arbitrary time instant $t$.

Combining this definition with Maxwell's equations, we could deduce the partial differential equation to govern the electromagnetic behavior of HTS subject to magnetic stimulation,

$$-\frac{1}{\mu_0}\left(\frac{\partial^2 A'_{sc,x}}{\partial y^2} + \frac{\partial^2 A'_{sc,x}}{\partial z^2}\right) + \sigma \frac{\partial A'_{sc,x}}{\partial t} + \sigma \frac{\partial A_{ex,x}}{\partial t} = 0. \tag{2}$$

where $A'_{sc,x}$ is the unknown to solve and the electric conductivity $\sigma$ of HTS is strongly dependent on the solution $A'_{sc,x}$ as well as the exotic excitation $A_{ex,x}$, which necessitates the action of numerical iteration.

The prominent advantage of this model is that, for the 2-D levitation problem concerned in this work, only the vector potential along the direction of translational invariance (the invisible $x$-axis in Fig. 1) needs to be defined and solved, which is rather profitable in terms of reducing the number of degrees of freedom and thus the computational time when adopting the finite-element technique to discretize the spatial domain including the HTS and the surrounding coolant as well. For detailed introduction of this method, one can refer to our previous publication in which this model was firstly proposed.[26]

The nonlinear feature of current–voltage relation in the HTS is characterized by a smoothed Bean–Kim's model of the critical state in the hyperbolic tangent approximation,[26,30]

$$J_x = J_{c0}\left(\frac{B_0}{|\mathbf{B}| + B_0}\right)\tanh\left(\frac{E_x}{E_0}\right) = J_{c0}\left(\frac{B_0}{|\nabla \times \mathbf{A}| + B_0}\right)\tanh\left(-\frac{\partial(A'_{sc,x} + A_{ex,x})}{\partial t}\frac{1}{E_0}\right). \tag{3}$$

where $J_{c0}$ is the critical current density in the absence of magnetic field, $E_0$ is the characteristic electric field and $B_0$ represents the critical magnetic flux density for which $J_c = J_{c0}/2$. Understanding the superconducting element is made of yttrium-barium cuprate cooled with liquid nitrogen, this work has used $E_0 = 5 \times 10^{-6}$ V/m and $B_0 = 0.25$ T.

## B. Numerical method

We follow the previous experience to numerically solve the partial differential equation of Eq. (2), i.e., the discretization of spatial domain using the finite-element technique is executed by the Galerkin's method, whereas the discretization of temporal domain is performed on the basis of the finite-difference technique via the backward Euler's scheme. The resultant nonlinear system of finite-element equation to resolve takes the form,[26]

$$\left( [\mathbf{K}(\mu_0)] + \frac{[\mathbf{Q}^n(\sigma)]}{\Delta t} \right) \{A'^n_{sc,x}\} = \frac{[\mathbf{Q}^n(\sigma)]}{\Delta t} \{\{A^{n-1}_{ex,x}\} - \{A^n_{ex,x}\} + \{A'^{n-1}_{sc,x}\}\}. \tag{4}$$

where $\Delta t$ is the time interval between the successive time instants, and the superscript $n$ and $n$-1 represent, respectively, the vector/matrix for the current and last time instant.

A linear triangular nodal element is deployed to generate the entries of the stiffness matrix $[\mathbf{K}(\mu_0)]$ and the damping matrix $[\mathbf{Q}^n(\sigma)]$. The vector $\{A_{ex,x}\}$ is known at each time instant and serves as the stimulated term from the PMG, which is calculated by an analytic method.[26,31] The Jacobian-free Newton-Krylov method[27] was applied to treat the nonlinearity of Eq. (4) with the associated algebraic equations after linearization solved by means of the generalized minimal residual algorithm.[32] This course avoids the usual evaluation of the Jacobian matrix for each element, and saves massive demands on computer memory and processing time thereby.

According to Lorentz's equation, the magnetic force per length exerting on the HTS along the $z$-direction, i.e., the levitation force $F_L$, can be numerically calculated by

$$F_L = \sum_{i=1}^{M} \iint_S J_x B_y dy dz = \sum_{i=1}^{M} \sum_{j=1}^{N} J^e_{x,j,i} B^e_{y,j,i} \Delta S^e_{j,i}. \tag{5}$$

where the superscript $e$ denotes the value of parameters at each element meshed by finite-element technique, and $\Delta S$ represents the area of each element, and $M$ is the amount of the HTS and $N$ the number of mesh element in each HTS. The magnetic flux density, $B_y$ and $B_z$, includes the contribution of the PMG as well as of the HTS.

## III. GENETIC ALGORITHM

The superconducting levitation problem is of highly nonlinear and non-differentiable, which causes its optimization to be computationally expensive even our theoretical model has been proven to be fast and robust.[25] In this situation, the extensive local optimization methods, though being easy to be implemented, are incompetent and the global optimizers should be considered. Among the global optimization algorithms at hand, the genetic algorithm is proven to be mature and excellent in the scientific community of electromagnetism and other disciplines.[33-35]

Being different from the local optimization methods, the optimization course of genetic algorithm was inspired by the Darwin's Theory of Evolution. The genetic algorithm defines three different classes, with the lowest one being chromosome, a concatenation of genes that represent a set of decision variables, and the middle one being individual that has merely one chromosome and respective operators, and the highest one being population which consists of a certain amount of individuals. Then genetic algorithm also defines operators to imitate the natural evolution, including the steps of selection, crossover and mutation. The optimal solution is searched in the genetic algorithm by manipulating a population of candidate solutions, and the best solutions are picked out by evaluating the fitness of all individuals in each population. The solutions with higher fitness will be selected as seeds to reproduce and crossover for creating a new generation. With the growth of the generation, the average fitness of the population will be improved gradually and become stable eventually when the evolutionary process of genetic algorithm comes to an end.

The objective function and the relevant constraint, which are dependent upon the studied problem, should be firstly defined to evaluate the fitness of each individual in the genetic algorithm. In this work, we search the maximum levitation force ($F_{L,max}$) subject to a constraint of the cross-sectional area ($S_{max}$) of the PMG, which is analogous to the case that minimizes the cross-section of the PMG to achieve the desired levitation force, as reported in Ref. 23. The task of the optimization in this work is to,

$$\text{Max.} \boldsymbol{Obj}(S) = F_{max}(S) - f_p(S), \text{ subject to } S \leq S_{max}, \tag{6}$$

where $f_p(S)$ is a penalty function defined as,

$$f_p(S) = \begin{cases} p(S - S_{max}), & S > S_{max} \\ 0, & S \leq S_{max} \end{cases}. \tag{7}$$

where $p$ is a positive constant severing as the penalty weight.

In addition, the fitness function is expressed as,

$$Fit(S) = \begin{cases} obj(S), & obj(S) \geq 0 \\ 0, & obj(S) < 0 \end{cases}. \tag{8}$$

This expression indicates that, the fitness of a set of design variables will be zero if its objective function is negative, viz., it has no probability to be inherited in the next generation.

Worthy of mention is that the guidance force, to ensure the lateral stability of the superconducting levitation, could be incorporated if the relevant constraint and the penalty function were defined and integrated in Eqs. (6)–(8). In this paper, only the optimization of levitation force will be released.

## IV. COMPUTATIONAL PERFORMANCE

The parameter selection of the genetic algorithm is important as it influences the computational time, which is generally expensive for the optimization of the nonlinear problem. In the following optimization, we use real-coded genes and random real number generators to produce the initial generation having a population of 50 individuals. Considering the practical condition of the PMG, the dimension of its magnet elements is restricted to be no more than 80 mm. The optimization will be terminated if the relative error between the adjacent generations is kept to be less than $10^{-4}$ for five successive generations, with a minimal amount of generation to prevent the prematuration. The possibilities to take the crossover and mutation are respectively 0.9 and 0.1.

With these parameters of genetic algorithm and setting $J_{c0} = 2.5 \times 10^8$ A/m$^2$, we tested the proposed optimization scheme upon the PMG in Fig. 1 to maximize the levitation force of HTS at a gap of 12 mm. The HTS is supposed to has geometry of 48 mm in width and 10 mm in thickness, whereas the cross-section of the PMG is constrained to be less than 7000 mm$^2$. The calculated development of the levitation force with the growth of the individual/generation was plotted in Fig. 2. This figure reveals that, the distribution of levitation force is rather scattered in the initial phase of the evolution where a global search takes place, and tends to be concentrative with the generation extended, although a few scattered point still exist due to the naturally required crossover and mutation of the genetic algorithm to make its individuals diverse. The levitation force of the best individual in each generation, connected as a line in Fig. 2, clearly displays that the optimization becomes converge and the optimum levitation force is constant with the growth of the generation. We conclude here that, the genetic algorithm program developed in this work has reproduced the reasonable evolution of the levitation force and could be used as a tool to optimize the superconducting levitation.

To further check the efficiency and robustness of the optimization scheme, we investigated the dependence of the optimum levitation force on the time step as well as on the number of mesh element in the finite-element calculation of the electromagnetic model, and the results were shown as a function of the constraints of cross-section area of PMG in Fig. 3.

The time step determines the temporal dimension of solving the diffusive partial differential equation to obtain the levitation force, and its choice influences the computational time, the numerical precision and stability. There is a tradeoff between the time cost and numerical precision in choosing the time step. Extreme choice of time step, wide or narrow, may both cause the diffusion of the numerical convergence. In Fig. 3, it shows a negligible difference among the optimum levitation force of the selected time steps, i.e., $\Delta t = 1$, 2 and 4 s, irrespective of the prescribed constraints of the cross-section. But, the computational time decreases sharply with the increase of the time step, which reduces the time cost of the optimization considerably. This finding demonstrates that, the developed optimization scheme is rather robust against the time step.

We selected three mesh element numbers of the spatial domain, from coarse to fine, to observe the influence of the number of mesh element on the optimization outcome, and the relevant results, plotted as an inset in Fig. 3, show that, the difference of the optimum levitation force at different constraints of the cross-section is not evident, allowing the flexible choice of the number of mesh element which mostly determines the computational time.

With the favorable findings abovementioned in reducing the computational time, we explore how efficient the developed optimization scheme could achieve with reasonable precision preserved. On an Intel Xeon E3-1230v3 processor-driven desktop running at the clock speed of 3.3 GHz, it was found that, the time cost of performing an optimization at a constraint of $S_{max} = 4000$ mm$^2$, is merely 3.6 hours with 3000 individuals evaluated if a multi-threaded preprocessor directives of OpenMP[36] was applied to the numerical program, which makes the best use of the native parallelism in the genetic algorithm. In this optimization, we set the time step to be 2 s and the number of mesh element in the superconducting and whole domain to be respectively 896 and 7560.

## V. CASE STUDIES

To provide the implications towards the practical optimization of the superconducting levitation, a set of case studies, being *inaccessible* from the experiments, were carried out to reveal how the working condition, geometrical and material characteristics of a HTS affect its maximum levitation force at different constraints of the cross-section of PMG, from 1000 to 7000 mm$^2$ with an interval of 1000.

### A. Working condition

Here the working condition refers to the desired levitation gap between the HTS and the PMG, which varies with different purposes. Supposing that the field-cooling height is 30 mm over the PMG and varying the levitation gap, we searched the maximum levitation force of a HTS, 48 mm in width and 10 mm in thickness, which could be achieved at different constraints of the cross-section of the PMG, and the results were plotted in Fig. 4.

It can be seen in Fig. 4, at a constraint of the cross-section of the PMG, the maximum levitation force will be always enhanced as the levitation gap decreases, which results in an improved cost-efficiency of the levitation system. At a given levitation gap, the maximum levitation force increases with the cross-section of the PMG, but the cost-efficiency, estimated by the slope of the curve, tends to be degraded especially for the cases of a small gap. These findings imply that, for a practical HTS levitation transit, the levitation gap should be minimized to improve its cost-efficiency, whereas the increase of the cross-section of the PMG will bring a decrease in the cost-efficiency though the levitation force could be accordingly heightened.

**B. Geometrical characteristics**

The geometrical characteristics of a 2-D HTS include the width and thickness, being respectively parallel to the lateral and vertical direction of the superconducting levitation system. With the field-cooling height and levitation gap kept to be respectively 30 and 12 mm, the influence of the geometrical characteristics of a HTS on its maximum levitation force at different constraints of the cross-section of the PMG was estimated by the proposed optimization scheme and, the relevant results were presented in Figs. 5 and 6 for the width and thickness respectively.

Figure 5 displays that, at a given constraint of $S_{max}$, the maximum levitation force of the HTS increases as its width grows, which become more evident with a higher $S_{max}$. This phenomenon indicates that a larger HTS in width could make better use of the increase in the cross-section of the PMG and thus owns a higher cost-efficiency, which also explains why the curve with the largest width, i.e., 60 mm, always has the highest slope. This observation confirms that, an optimum relation between the width of the HTS and the cross-section of the PMG really exists, as reported in the previous work.[9,11,12,15,20]

Figure 6 displays that, at the smallest value of $S_{max}$, the maximum levitation force is nearly independent of the thickness of the HTS, demonstrating that the HTS is partially excited to generate the levitation force as the applied field of the PMG with a small cross-section is weak. The curves for different thickness then begin to separate from each other and this separation becomes more and more remarkable as the value of $S_{max}$ rises, revealing an increasing portion of HTS is excited to upgrade the levitation force. Another useful finding for application in this study is that, the maximum levitation force obtained at the smallest thickness tends to be insensitive to the cross-section of PMG. In this view, the HTS has been nearly fully excited by the applied field of PMG and to enhance the levitation force by enlarging the PMG is totally uneconomical for such HTS. It is concluded that, the optimum thickness of HTS is strongly related with the constraint of the cross-section of the PMG, and a thick HTS could provide much higher levitation force than a thin one only if the cross-section of the PMG is sufficient.

**C. Critical current density**

The material performance to evaluate the HTS for levitation purpose is generally represented by the critical current density. With the working condition and the geometry unvaried, we appraised the role of the critical current density in improving the maximum levitation force of a HTS at different constraints of the cross-section of the PMG, and the results were presented in Fig. 7. The selected values of critical current density, i.e., $1\times 10^8$, $2.5\times 10^8$, and $5\times 10^8$ A/m$^2$, refer to respectively the HTS with a poor, fair and excellent performance of superconductivity.[6]

It is clearly illustrated in Fig. 7 that, the promotion of the critical current density to be fair could bring out a significant enhancement of the maximum levitation force, especially for the large cross-section of the PMG, where higher performance of HTS will make better use of the applied field of the PMG and thus provides a higher levitation force. This figure also indicates that, the maximum levitation force obtained by the worst HTS becomes saturated rapidly with the cross-section of the PMG, and by contrast, it increases continuously for the best HTS even with a decreased slope. Moreover, when the cross-section of the PMG is constrained to be 2000 mm$^2$ or less, the maximum levitation force obtained by the best HTS has no clear improvement as compared to that of the fair one. As the cost of a HTS is mostly determined by its performance, we recommend using a poor HTS to assemble the maglev system if the cross-section of the PMG is limited, whereas using a fair or excellent HTS to exploit the applied field at most if a large cross-section of the PMG is permitted.

## VI. CONCLUSIONS

With the HTS electromagnetically modeled by the quasistatic approximation of a generalized vector potential approach and numerically solved by the finite-element technique, an efficient scheme for optimizing the HTS-PMG maglev transit has been proposed and developed by the use of the genetic algorithm to maximize the levitation force at the constraint of the cross-section of the PMG. Considering a typical task of optimization of the HTS-PMG maglev system, the proposed scheme was found to be very robust against the coarseness of the mesh and the time step, which makes the time cost be merely 3.6 hours with 3000 individuals evaluated on a moderate desktop with the nonlinear behavior of the HTS included.

Taking advantage of the proposed scheme, we performed a set of case studies from a practical point of view to understand how the working condition, geometrical and material characteristics of a HTS affect its maximum levitation force over a PMG with different constraints of the cross-section. These studies suggest that, irrespective of the constraint of the PMG, the cost-efficiency of the system could be considerably improved through minimizing the levitation gap and enlarging the width of the HTS. However, the increase in the thickness or the critical current of the HTS seems to be effective to improve the cost-efficiency only if the cross-section of the PMG is sufficient. The benefit from enhancing the geometrical and material parameters of the HTS to increase the levitation force becomes elevated as the cross-section of the PMG grows. These physical findings are logically reasonable and enable the proposed scheme to be applicable to optimize the superconducting levitation for practical design.

## ACKNOWLEDGEMENTS

This work was supported in part by the National Natural Science Foundation of China (Grant No. 51475389), and by the Fundamental Research Funds for the Central Universities (Grant No. 2682014CX039), and by the Self-determined Projects of the State Key Laboratory of Traction Power (Grants No. 2013TPL_T05 and 2015TPL_T05).

**Figures**

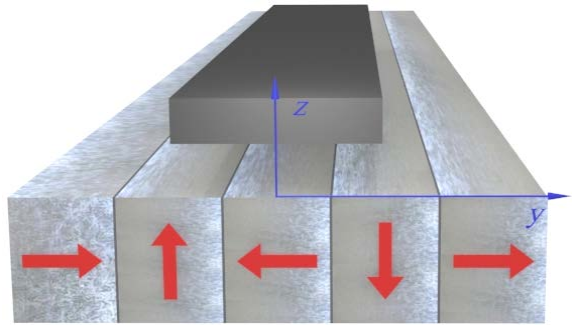

FIG. 1. Geometrical configuration of a Halbach-derived permanent magnet guideway (PMG) with a high temperature superconductor levitated above, both being invariant along the invisible *x*-direction and symmetric in terms of the *z*-direction of a Cartesian coordinate system *x*, *y*, *z*. A magnetization of $8.753 \times 10^5$ A/m was assigned to all magnet elements, intending to approximately reproduce the performance of an N35 magnet.

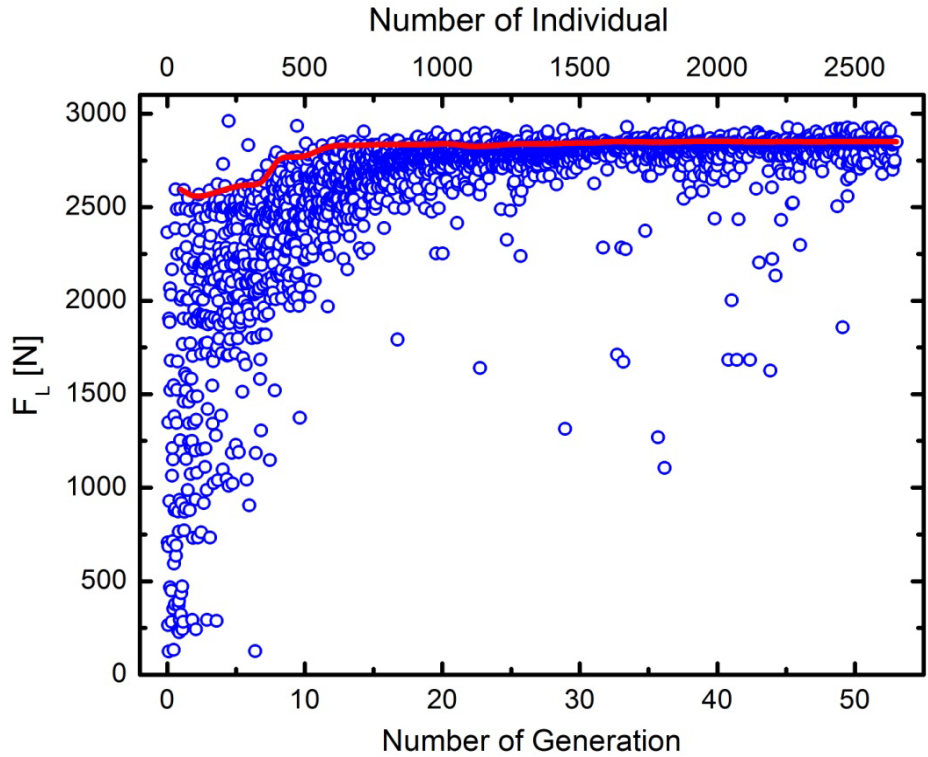

FIG. 2. A representative development of the levitation force ($F_L$) as the individual/generation evolves while using the genetic algorithm to optimize the superconducting levitation system. The open circle denotes the achieved levitation force at each individual, of which the best one in each generation was plotted (solid line) to display the trend of the optimum levitation force in the process of the evolution. In this calculation, the superconducting bulk, with geometry of 48 mm in width and 10 mm in thickness, was field-cooled at a position of 30 mm over the PMG and the position where the levitation force was optimized was 12 mm over the PMG. The cross-section of the PMG was constrained to be less than 7000 mm$^2$ in this calculation.

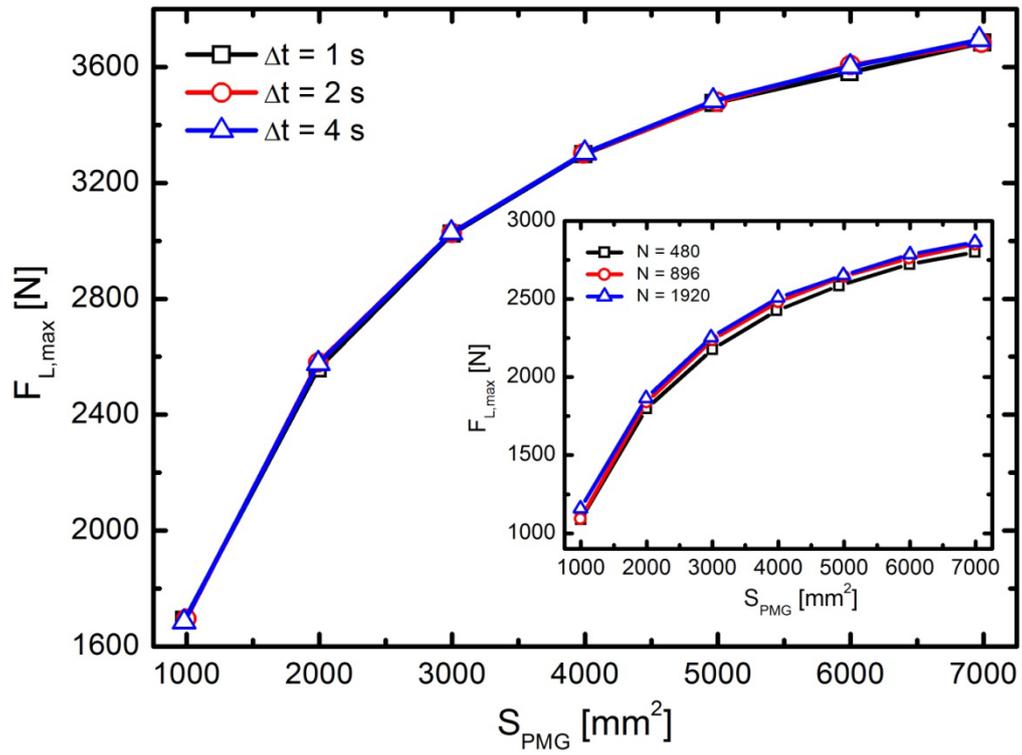

FIG. 3. Illustration of the influence of the time step $\Delta t$ as well as the number of mesh element $N$ in the superconducting domain (inset) on the maximum levitation force ($F_{L,max}$) at different constraints upon the cross-section of the PMG ($S_{PMG}$). Selected time step shown in this figure is $\Delta t = 1$, 2 and 4 s with $N = 896$, whereas the selected number of mesh element in the superconducting domain is $N = 480$, 896 and 1920 with $\Delta t = 2$ s.

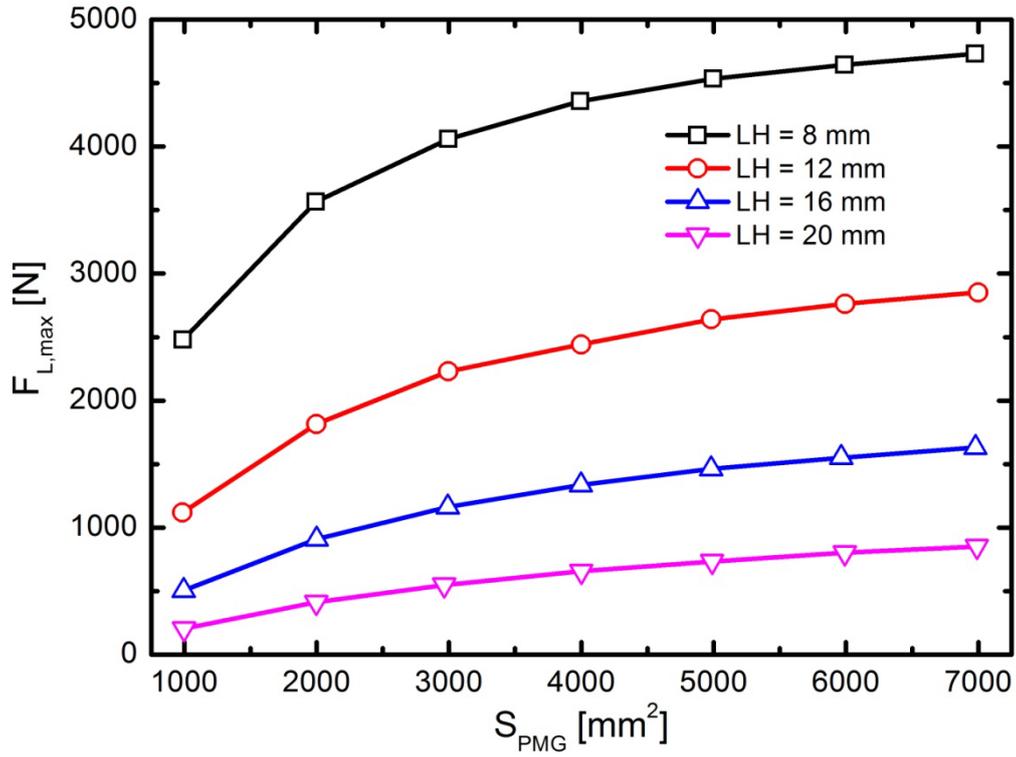

FIG. 4. The maximum levitation force ($F_{L,max}$) as a function of the constraint upon the cross-section of the PMG ($S_{PMG}$) at different levitation heights (LH) where the levitation force was optimized. In this calculation, the HTS, with 48 mm in width and 10 mm in thickness, was field-cooled at a position of 30 mm above each PMG. It was assumed that $J_{c0} = 2.5 \times 10^8$ A/m$^2$ in this calculation.

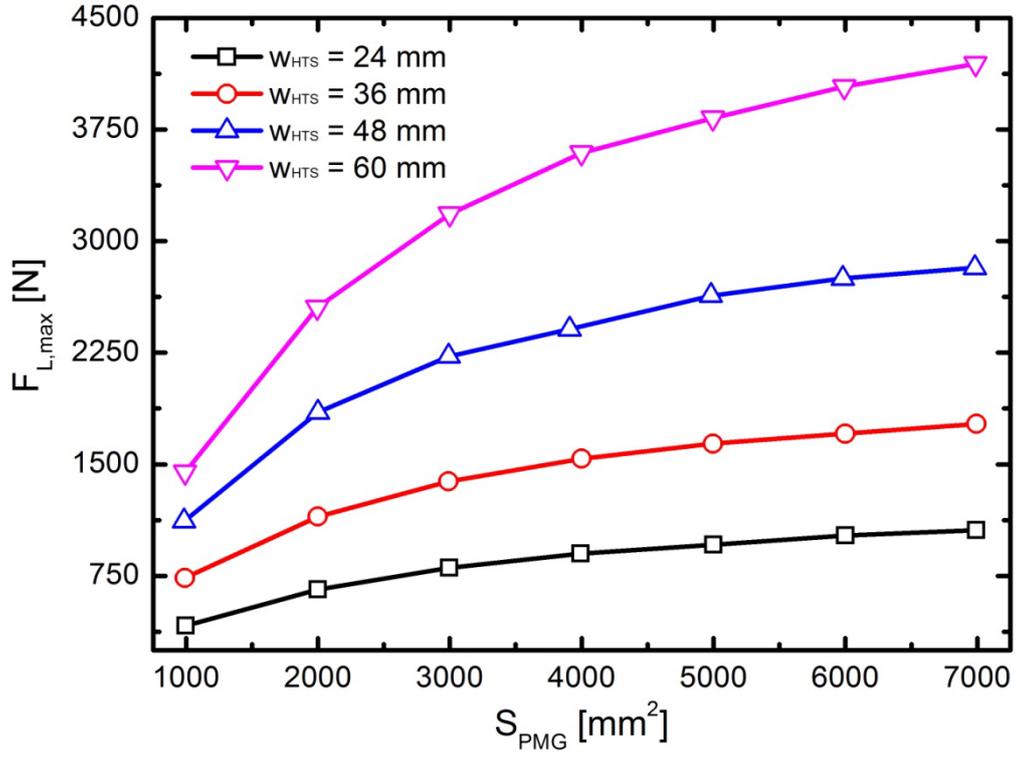

FIG. 5. The maximum levitation force ($F_{L,max}$) as a function of the constraint upon the cross-section of the PMG ($S_{PMG}$) with a varied width ($w_{HTS}$) of the HTS. In this calculation, the HTS was field-cooled at a position of 30 mm and its levitation force was optimized at a position of 12 mm over the PMG. It was assumed that $J_{c0} = 2.5 \times 10^8$ A/m$^2$ and the thickness is 10 mm in this calculation.

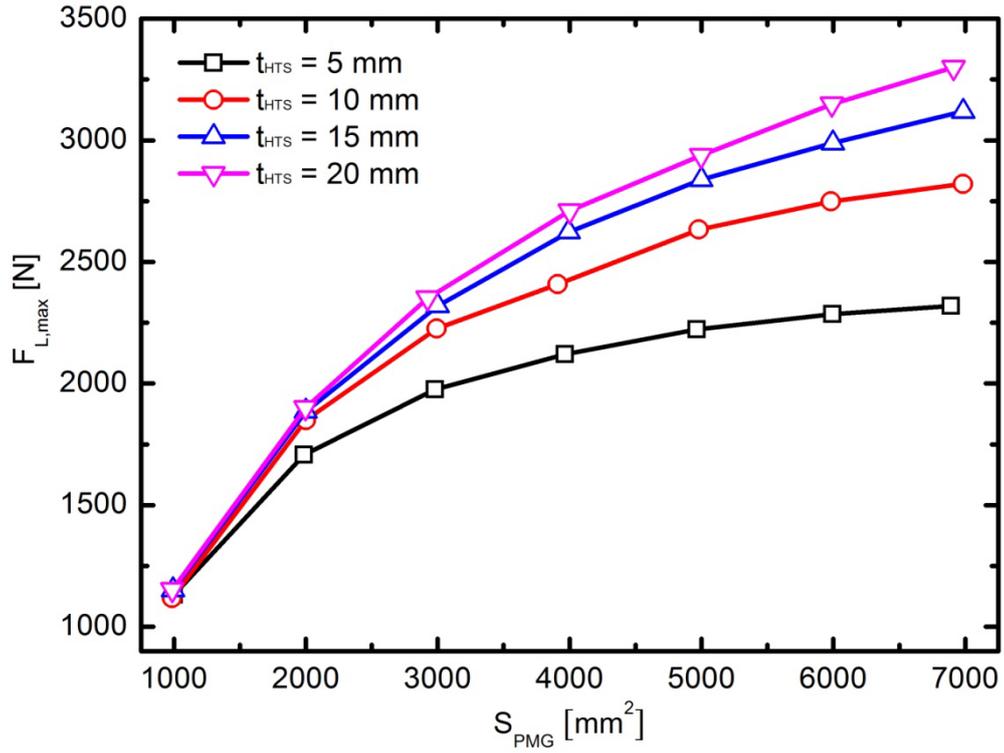

FIG. 6. The maximum levitation force ($F_{L,max}$) as a function of the constraint upon the cross-section of the PMG ($S_{PMG}$) with a varied thickness ($t_{HTS}$) of the HTS. In this calculation, the HTS, with an unvaried width of 48 mm, was field-cooled at a position of 30 mm and its levitation force was optimized at a position of 12 mm over the PMG. It was assumed that $J_{c0} = 2.5 \times 10^8$ A/m$^2$ and the width is 48 mm in this calculation.

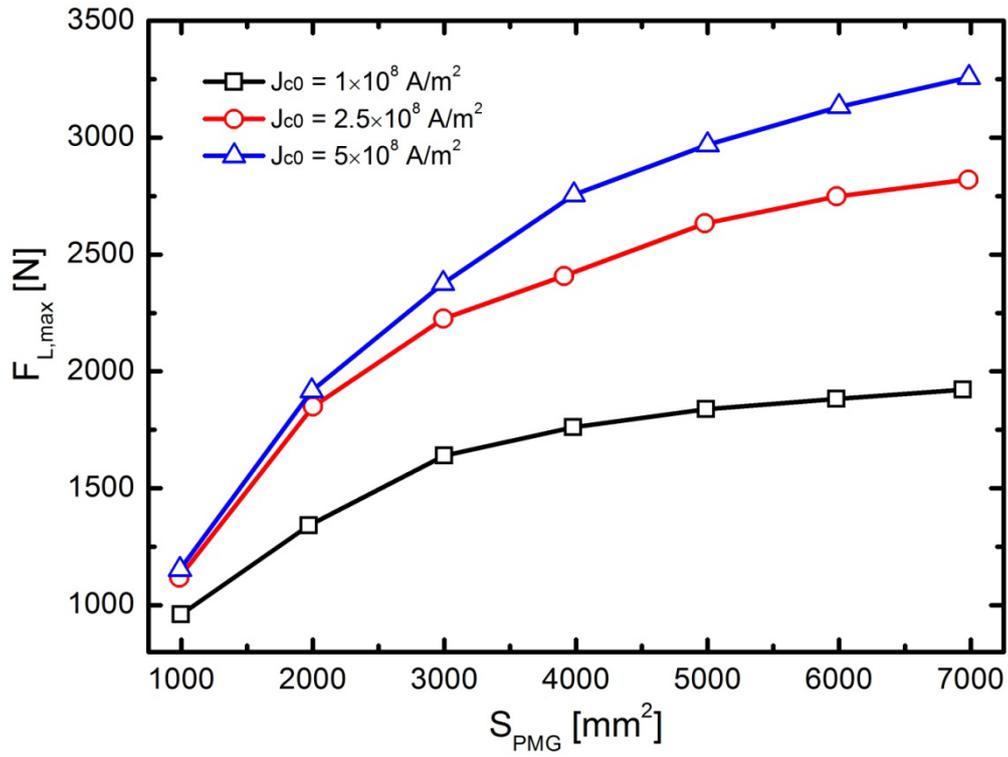

FIG. 7. Illustration of the influence of the critical current density on the maximum levitation force ($F_{L,max}$) under different constraints upon the cross-section of the PMG ($S_{PMG}$). In this calculation, the HTS, with 48 mm in width and 10 mm in thickness, was field-cooled at a position of 30 mm and its levitation force was optimized at a position of 12 mm over the PMG.

Press, Cambridge, 2008).